 \documentclass[twocolumn,prl,aps,psfig,preprintnumbers,superscriptaddress]{revtex4-2}
\usepackage{amssymb}
\usepackage{amsmath}
\usepackage{graphicx}
\usepackage{dcolumn}
\usepackage{bm}
\usepackage{color}
\usepackage{epstopdf}
\begin{document}

\preprint{}

\title{Charge disproportionation in the spin-liquid candidate $\kappa$-(ET)$_2$Cu$_2$(CN)$_3$ at $6$~K revealed by $^{63}$Cu NQR measurements}

\author{T. Kobayashi}
 \email{tkobayashi@phy.saitama-u.ac.jp}
 \affiliation{Graduate School of Science and Engineering, Saitama University, Saitama, 338-8570, Japan}
 \affiliation{Ames Laboratory, U.S. DOE, and Department of Physics and Astronomy, Iowa State University, Ames, Iowa 50011, USA}
\author{Q.-P. Ding} 
\affiliation{Ames Laboratory, U.S. DOE, and Department of Physics and Astronomy, Iowa State University, Ames, Iowa 50011, USA}
\author{H. Taniguchi}
\author{K. Satoh}
 \affiliation{Graduate School of Science and Engineering, Saitama University, Saitama, 338-8570, Japan}
\author{A. Kawamoto}
\affiliation{Department of Condensed Matter Physics, Graduate School of Science, Hokkaido University, Sapporo 060-0810, Japan}
\author{Y. Furukawa}
 \affiliation{Ames Laboratory, U.S. DOE, and Department of Physics and Astronomy, Iowa State University, Ames, Iowa 50011, USA}

\date{\today}

\begin{abstract}
The spin-liquid candidate $\kappa$-(ET)$_2$Cu$_2$(CN)$_3$ [ET: bis(ethylenedithio)tetrathiafulvalene] does not exhibit magnetic ordering down to a very low temperature, but shows a mysterious anomaly at $6$~K.
 The origin of the so-called $6$~K anomaly is still under debate. 
 We carried out nuclear quadrupole resonance (NQR) measurements on the copper sites of the insulating layers, which are sensitive to the charge dynamics unlike the conventional spin-$1/2$ nuclear magnetic resonance (NMR). 
The main finding of this study is that the observation of a sharp peak behavior in the nuclear spin-lattice relaxation rate $T_1^{-1}$ of $^{63}$Cu NQR at $6$~K while $T_1^{-1}$ of both $^{13}$C and $^{1}$H NMR show no clear anomaly. 
This behavior can be understood as a second-order phase transition related to charge disproportionation in the ET layers. 
\end{abstract}

\maketitle

Quasi-two-dimensional organic charge transfer salts $\kappa$-(ET)$_2X$ [ET and $X$ denote bis(ethylenedithio)tetrathiafulvalene and monovalent anion, respectively] possess half-filled band owing to strong dimeric structures of donor molecules.
In the typical phase diagram, antiferromagnetic and superconducting phases are adjacent to each other  and the phase transition between them has been discussed in terms of the bandwidth and on-site Coulomb repulsion \cite{Kanoda1997c}, where the magnetic properties mainly originate from  antiferromagnetic interaction between (ET)$_2^+$ dimers with $S$=$1/2$.
The role of the charge degree of freedom has also been pointed out to explore the recent observations of dielectric anomaly with antiferromagnetic ordering \cite{Lunkenheimer2012}, charge order \cite{Drichko2014,Gati2018a}, and quantum dipole liquid \cite{Hassan2018}.

Such complex physical properties related to spin and charge degrees of freedom have also been discussed in a representative spin-liquid candidate $\kappa$-(ET)$_2$Cu$_2$(CN)$_3$. 
This material does not show long range magnetic ordering down to $T$=$32$~mK despite a large antiferromagnetic exchange interaction of $\sim 250$~K \cite{Shimizu2003}, and therefore it has attracted much attention as the quantum spin liquid. 
$\kappa$-(ET)$_2$Cu$_2$(CN)$_3$ showed a relaxor-like dielectric response below $60$~K, and intradimer charge disproportionation (CD) was proposed \cite{Abdel-Jawad2010} although the mechanism is under debate \cite{Sedlmeier2012,Pinteric2015,Dressel2016}. 
 It is also pointed out that the mysterious anomaly at $6$~K could be related to charge properties.
This anomaly was initially detected by the hump structure in the $T$ dependence of specific heat whereby a crossover from a thermal disordered to a quantum spin liquid state was suggested \cite{Yamashita2008}. 
On the other hand, the thermal expansion measurements indicate a phase transition, and the importance of charge degree of freedom was suggested \cite{Manna2010}. 
 Whereas many experimental and theoretical efforts have been devoted to elucidating the origin of the so-called \lq\lq{6~K anomaly\rq\rq{}} \cite{Pratt2011,Poirier2012,Itoh2013,Poirier2014,Isono2016,Furukawa2018, Kawamura1984,Baskaran1989,Lee2007,Qi2008,Grover2010,Riedl2019}, it is not even clear whether it is a phase transition or a crossover phenomenon.

 Nuclear magnetic resonance (NMR) is an effective technique for providing the microscopic evidences of charge and/or spin anomalies. 
  Up to now, there have been several NMR reports using $^{13}$C and $^1$H nuclei with nuclear spin $I$=$1/2$. 
  However, such NMR measurements cannot probe the charge directly because of no direct interaction between $I$=$1/2$ nucleus and charge, whereas one can obtain the information of the charge distributions through the change in hyperfine coupling constants \cite{Miyagawa2000}. 
 Therefore NMR measurements using nuclei with $I$=$1/2$ are less sensitive to charge anomaly. 
 In fact, no clear anomaly at $6$~K has been observed in $^1$H- and $^{13}$C-NMR measurements, especially, in nuclear spin-lattice relaxation measurements \cite{Shimizu2003,Kawamoto2004a,Shimizu2006,Saito2018}.
  Here, we focus on the nuclear quadrupole resonance (NQR) technique, which is a charge-sensitive probe offering a new perspective. 
This is because the NQR technique directly detects the electric field gradient (EFG) through the nuclear quadrupole moment $Q$ ($\neq 0$ when $I>1/2$). 
  Until now, NQR method has attracted less attention in the field of organic conductors since there is no NQR-active nuclei in the ET molecule. 
  In this Letter, we report the first NQR experiment in the ET-based charge transfer salts using copper nuclei located at the insulating layer in $\kappa$-(ET)$_2$Cu$_2$(CN)$_3$.
 Our NQR results elucidate that the $6$~K anomaly can be understood as a phase transition related to the charge degree of freedom in the molecular layers. 

Polycrystalline samples were prepared by the standard electrochemical reaction \cite{Geiser1991a}. 
Zero-magnetic-field (ZF) NQR experiments of $^{63}$Cu ($I$=$3/2$, $Q$=$-0.21$ barns) and $^{65}$Cu ($I$=$3/2$, $Q$=$-0.195$ barns) were performed by using a home-made phase-coherent spin-echo pulse spectrometer. 
$^{63,65}$Cu-NQR spectra were obtained in steps of frequency by measuring the intensity of the Hahn spin echo. 
The nuclear spin-lattice relaxation rate of $^{63}$Cu, $^{63}T_1^{-1}$ was measured with a saturation recovery method and determined by fitting $M(t)$ using the stretched exponential function $1-M(t)/M(\infty)=\exp[-(3t/T_1)^{\beta}]$, where $M(t)$ and $M(\infty)$ are the nuclear magnetization at time $t$ after the saturation and the equilibrium nuclear magnetization at $t \to \infty$, and $\beta$ is the stretched exponent, respectively. 
 The nuclear spin-echo decay rate $T_2^{-1}$ was determined by the spin-echo signal $M(2\tau)$ as a function of time $2\tau$, where $\tau$ is the interval between the first exciting and refocusing pulses. 

%%%%%%%%%%%%%%%%%%%%%%%%%%%% FIG1 %%%%%%%%%%%%%%%%%%%%%%%%%%%%%%%%%%%
\begin{figure}[tb]
\includegraphics[width=\columnwidth]{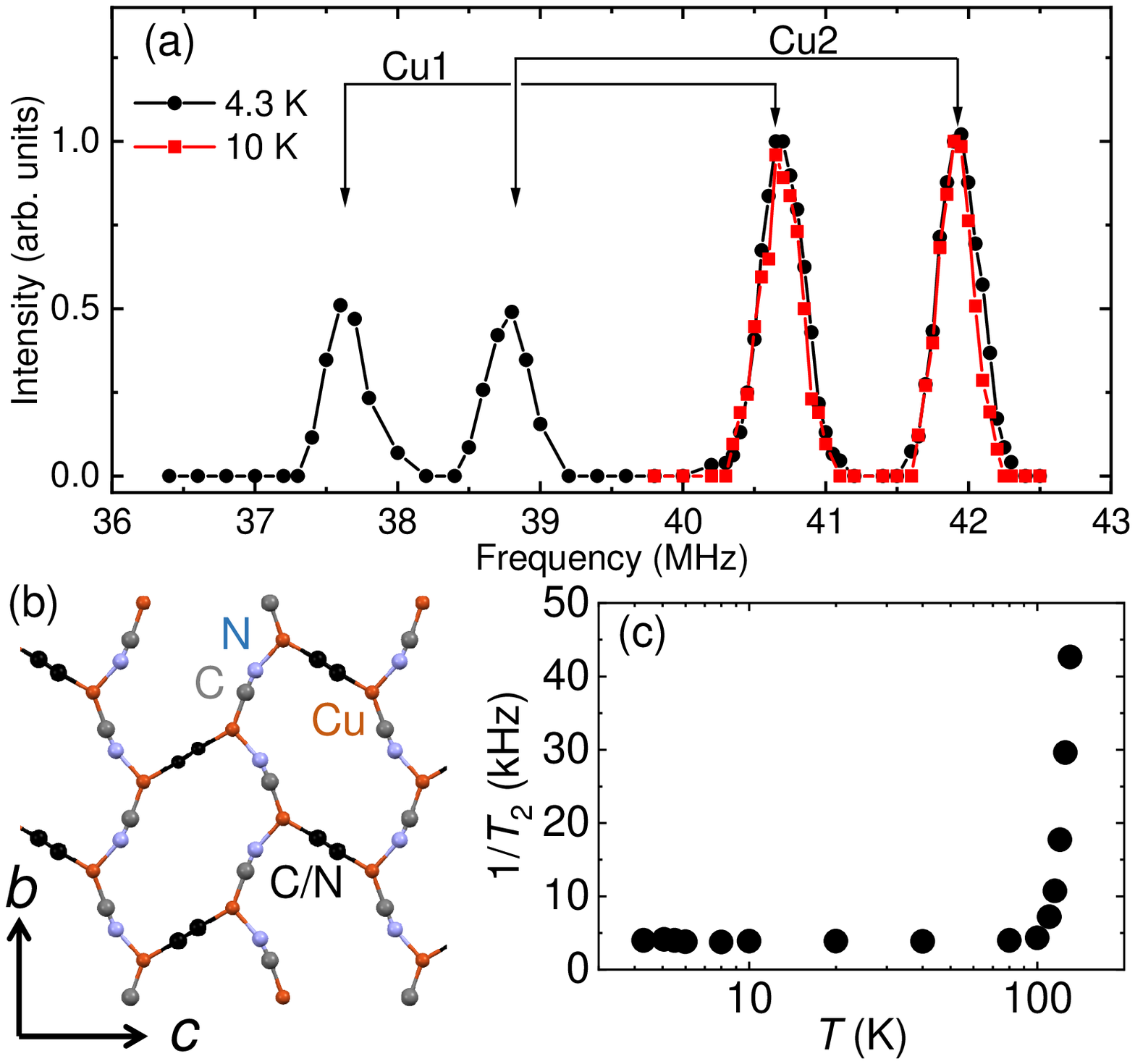} 
\caption{
(a) ZF NQR spectra at $4.3$~K (black circles) and $10$~K (red squares).
(b) Structure of the Cu$_2$(CN)$_3$ insulating layer. 
(c) $T$ dependence of $T_2^{-1}$ measured at the peak around 40.7 MHz.
}
\label{fig1}
\end{figure}
%%%%%%%%%%%%%%%%%%%%%%%%%%%% FIG1 %%%%%%%%%%%%%%%%%%%%%%%%%%%%%%%%%%%

 Figure~\ref{fig1}(a) shows the NQR spectra at $4.3$~K where four lines are observed in the frequency range of $36$-$43$~MHz. 
 In the case of $I$=$3/2$, one expects a single NQR line at the resonance frequency of $\nu_{\rm NQR}$=$\nu_{\rm Q}\sqrt{1+\eta^2/3}$,
where $\nu_{\rm Q}$ is nuclear quadrupole frequency defined by $\nu_{\rm{Q}}$=$eQV_{ZZ}$/2$h$.
 Here $e$, $V_{ZZ}$, $h$, and $\eta$ are elementary charge, the principal value of the EFG tensor, Planck's constant, and asymmetry parameter of EFG, respectively. 
  Since there are two isotopes $^{63}$Cu and $^{65}$Cu, NQR spectrum must be a pair of lines with different intensities where the $^{63}$Cu-NQR intensity is about double of the $^{65}$Cu-NQR one due to the natural abundances of the two nuclei (viz, $^{63}$Cu: 69\%; $^{65}$Cu: 31\%). 
  Therefore, the observation of four lines clearly indicates the existence of two Cu sites with slightly different environments. 
  By taking the difference in $Q$ between $^{63}$Cu and $^{65}$Cu into consideration, we can assign the two pairs (defined by Cu1 and Cu2 sites) as shown in Fig.~\ref{fig1}(a).

 In $\kappa$-(ET)$_2$Cu$_2$(CN)$_3$, as shown in Fig.~\ref{fig1}(b), there are three cyano groups around a Cu nucleus, and one of them is considered to be positionally disordered with a $50$~\% carbon and $50$~\% nitrogen distribution (depicted by black symbols) due to an inversion point at the center of the cyano groups \cite{Jeschke2012,Foury-Leylekian2018}. 
 Consequently, Cu$^{+}$ ions are expected to be trigonally coordinated with two carbons and one nitrogen, or with one carbon and two nitrogens, making two copper sites with slightly different environments (${\it i.e.}$, slightly different EFG) with equal probability.
 This is consistent with the observed NQR spectrum including the intensity ratio of 1:1 for Cu1 and Cu2.
 Although we do not know which one is which, our observation directly evidences the disorder of carbon and nitrogen ions due to the inversion symmetry in $\kappa$-(ET)$_2$Cu$_2$(CN)$_3$. 

 It is noted that one cannot determine the values of $\nu_{\rm{Q}}$ and $\eta$ for the Cu ions separately from only the NQR spectrum. 
 A finite value of $\eta$, the lack of axial symmetry of EFG, is expected from the local symmetry at the Cu sites.  
 With the help of NMR spectrum measurements described below, $\eta$ is estimated to be $\sim0.5$, and thus the values of $\nu_{\rm Q}$ are estimated to be $36.2$($37.3$)~MHz and $39.1$($40.3$)~MHz for $^{65}$Cu and $^{63}$Cu, respectively, for the Cu1(Cu2) sites.
 The direction of $V_{ZZ}$ is considered to be parallel to the $a^*$ axis, perpendicular to the insulating layers.

  As shown in Fig.~\ref{fig1}(a), there is no significant difference in the spectra between $4.3$~K and $10$~K for the two Cu sites, hereafter, we show the $T$ dependence of spectra and $^{63}T_1^{-1}$ measured at the peak at around $40.7$~MHz ($^{63}$Cu ions at the Cu1 site).
  NQR experiments were able to be conducted between $1.5$-$120$~K. 
  Above $100$~K, $T_2^{-1}$ suddenly increases as shown in Fig.~\ref{fig1}(c), and eventually the NQR signals disappeared above $130$~K owing to the shortening of $T_2$. 
 This is probably due to the vibrational motion of ethylene end groups of ET molecule at high $T$ \cite{Kurosaki2005}.

%%%%%%%%%%%%%%%%%%%%%%%%%%%% FIG2 %%%%%%%%%%%%%%%%%%%%%%%%%%%%%%%%%%%
\begin{figure}[tbp]
\begin{center}
\includegraphics[width=\columnwidth]{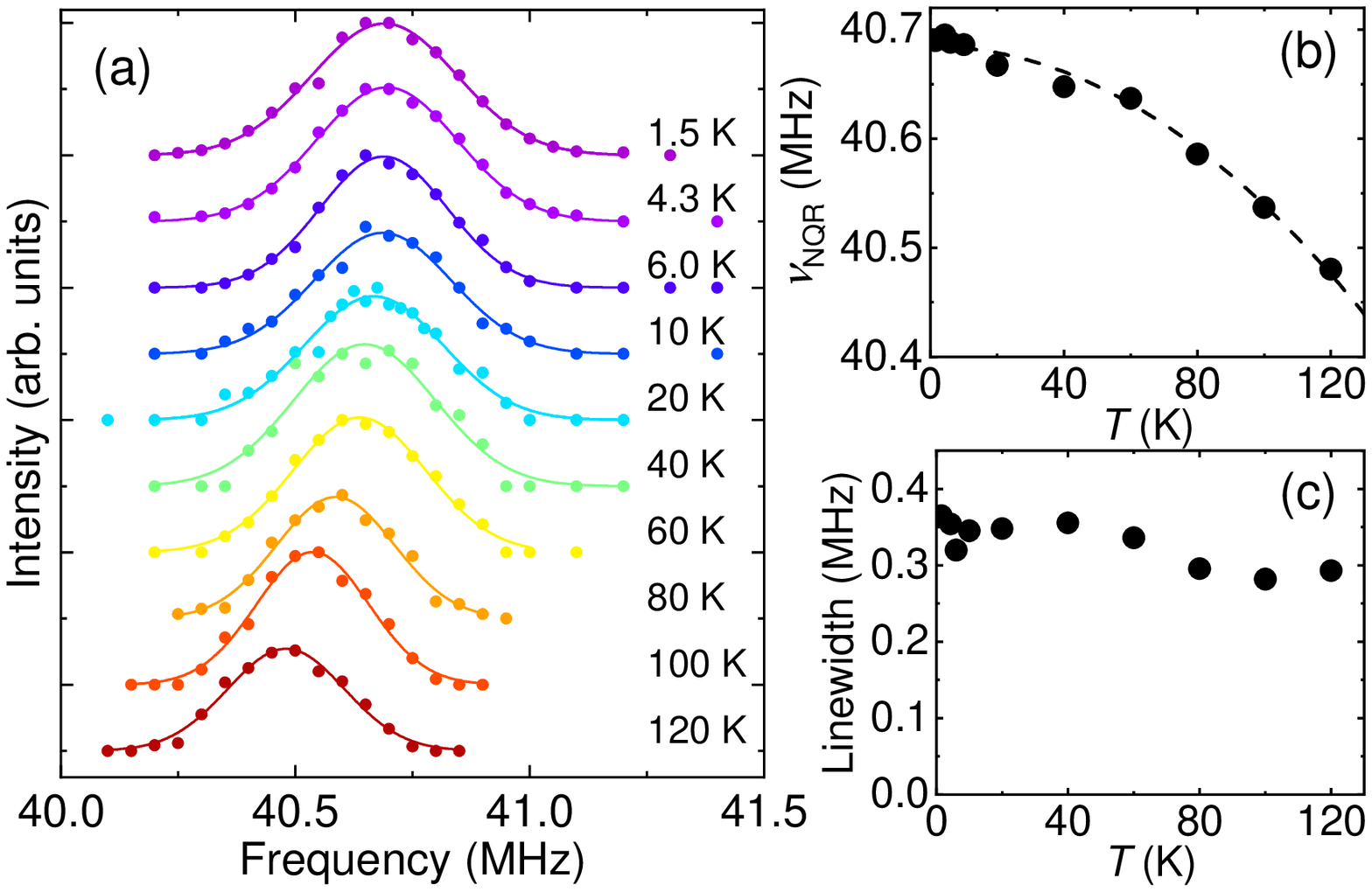}
\end{center}
\caption{
(a) $T$ evolution of NQR spectra. Solid lines are the fitting curves using Gaussian function.
(b) $T$ dependence of $\nu_{\rm NQR}$. 
The broken curve is the calculated result with the empirical formula $\nu_{\rm NQR} = \nu_0 \exp(-\alpha T^{2})$ (see text).
(c) $T$ dependence of linewidth. 
}
\label{fig2}
\end{figure}
%%%%%%%%%%%%%%%%%%%%%%%%%%%% FIG2 %%%%%%%%%%%%%%%%%%%%%%%%%%%%%%%%%%%

Figure~\ref{fig2}(a) shows the $T$ evolution of $^{63}$Cu-NQR spectra for the Cu1 site. 
There is no drastic change in the NQR spectra suggesting no structural phase transition in the $T$ range.
The peak position is slightly shifted to lower frequency with increasing $T$, corresponding to the decrease in $\nu_{\rm NQR}$ as shown in Fig.~\ref{fig2}(b). 
The $T$ dependence of $\nu_{\rm NQR}$ is considered to be originated from thermal lattice expansion and can be described by the empirical formula \cite{Koukoulas1990,Iwase2007}, $\nu_{\rm NQR}$=$\nu_0 \exp(-\alpha T^{2} $).
As shown by the curve in Fig.~\ref{fig2}(b), the $T$ dependence of $\nu_{\rm NQR}$ is well reproduced with the formula with $\nu_0$=$40.68$~MHz and $\alpha$=$3.58 \times 10^{-7}$~K$^{-2}$.
Figure~\ref{fig2}(c) shows the $T$ dependence of linewidth (full width at half maximum, FWHM) determined by the fitting of the spectra with Gaussian function.
With decreasing $T$, the linewidth increases from $0.3$~MHz at $120$~K to $0.36$~MHz at around $60$~K, and is nearly independent of $T$ below $60$~K.

\begin{figure}[tb]
\begin{center}
\includegraphics[width=\columnwidth]{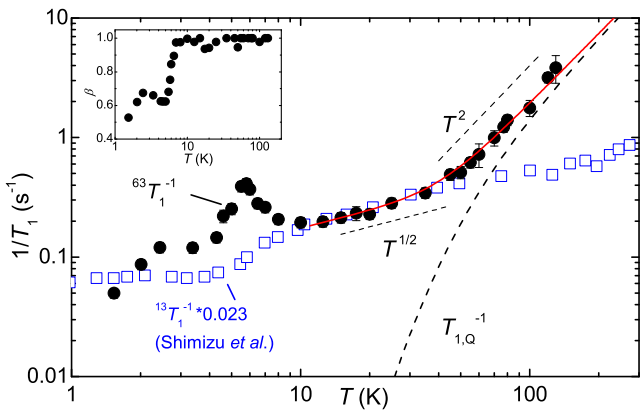}
\end{center}
\caption{
$T$ dependencies of $^{63}T_1^{-1}$ (black circles) and $^{13}T_1^{-1}$ (open blue squares) from Ref. \cite{Shimizu2006}.
The values of $^{13}T_1^{-1}$ are reduced by a factor of $0.023$. 
Dashed black and solid red curves represent $T_{1,Q}^{-1}$ due to the thermal lattice vibration, and  $T_{1,Q}^{-1}+0.055T^{1/2}$, respectively. 
Inset shows the $T$ dependence of $\beta$.}
\label{fig3}
\end{figure}

 Figure~\ref{fig3} shows the $T$ dependence of ZF $^{63}T_1^{-1}$. 
 For comparison, the $T$ dependence of $T_1^{-1}$ of $^{13}$C, $^{13}T_1^{-1}$, is also plotted \cite{Shimizu2006}.
 Note that the values of $^{13}T_1^{-1}$ are reduced by a factor of $0.023$. 
The $T$ dependences  are quite different from each other. 
 $^{63}T_1^{-1}$ shows $T^2$ dependence above $60$~K, below which it is proportional to $T^{1/2}$ coinciding with that of $^{13}T_1^{-1}$. 
 With further decreasing $T$ below $10$~K, $^{63}T_1^{-1}$ starts to increase and then shows a pronounced peak at $6$~K, whereas $^{13}T_1^{-1}$ decreases monotonically. 
  The exponent $\beta$ deviates from unity below $6$~K as shown in the inset of Fig.~\ref{fig3}, indicating a development of inhomogeneity in $^{63}T_1$. 
A similar inhomogeneity below $\sim$ $6$~K was also observed in $^{13}$C- and $^1$H-NMR measurements \cite{Shimizu2003,Shimizu2006}.

%%%%%%%%%%%%%%%%%%%%%%%%%%%% FIG3 %%%%%%%%%%%%%%%%%%%%%%%%%%%%%%%%%%%
%%%%%%%%%%%%%%%%%%%%%%%%%%%% FIG3 %%%%%%%%%%%%%%%%%%%%%%%%%%%%%%%%%%%

The $T^{\sim2}$ dependence of $^{63}T_1^{-1}$ observed above $60$~K can be explained by the thermal vibrations of the three nearest neighbor CN$^{-}$ ions with respect to Cu ion.
  A similar $T$ dependence has been reported in other kinds of diamagnetic insulators \cite{Nakayama1982,Iwase2007,Iwase2010}, whose $T$ dependence was explained by the quadrupolar relaxation due to the two-phonon Raman process \cite{Abragam1961}. 
  In this case, the quadruple relaxation ($T_{1,{\rm Q}}^{-1}$) is described as 
$ T_{1,{\rm Q}}^{-1}$=$\frac{81 \pi}{2} \left( \frac{F_2 \hbar}{mv^2}\right)^2 \int_{0}^{\Omega} e^{\hbar \omega/k_B T} (e^{\hbar \omega/k_B T}-1)^{-2}(\omega/\Omega)^6 d\omega$. 
 Here, $m$, $v$, and $\Omega$ are the atomic mass of $^{63}$Cu, sound velocity in the crystal, and cutoff frequency related to Debye temperature ($\Theta$), respectively. 
 $F_2$ is a parameter which can be approximated by $2\pi \nu_Q$ \cite{Abragam1961,Klanjsek2017}. 
 Assuming $v$=$10^3$~m/s \cite{ImajoSound}, and a typical value $\Theta=\hbar \Omega/k_B = 200$~K for $\kappa$-(ET)$_2X$ \cite{Nakazawa1997,Andraka1989}, we have calculated the $T$ dependence of $T_{1,{\rm Q}}^{-1}$ . 
The dashed black curve is the calculated result without free parameter, which reproduces the experimental data very well by adding another contribution as described below.

 From the calculated result of $T_{1,{\rm Q}}^{-1}$, $^{63}T_1^{-1}$ is expected to decrease drastically at low $T$ due to the suppression of thermal vibrations. 
 However, $^{63}T_1^{-1}$ gradually deviates from $T_{1,{\rm Q}}^{-1}$ below $\sim 60$~K and shows $T^{1/2}$ dependence between $10$-$40$~K, which is the same as that of $^{13}T_{1}^{-1}$ \cite{Shimizu2006}.
 Since $^{13}T_{1}^{-1}$ originates from the magnetic fluctuations from the $\pi$ electrons of the ET layers \cite{Kawamoto2006a,Shimizu2006}, the similar $T$ dependence of $^{63}T_1^{-1}$ indicates that the magnetic fluctuations become dominant at the Cu site. 
 In fact, this interpretation can be confirmed by looking at the ratio of $^{63}T_1^{-1} / ^{13}T_1^{-1}$.
Since $T_1^{-1}$ is proportional to the square of gyromagnetic ratio $\gamma$ and hyperfine coupling constant $A$, the ratio of $(^{63}\gamma^{2}~ ^{63}T_1)^{-1} / (^{13}\gamma^{2}~ ^{13}T_1)^{-1} \sim 0.02$ should be proportional to $|^{63}A$/$^{13}A|^2$. 
To estimate the $^{63}A$, we performed $^{63}$Cu-NQR measurements on the antiferromagnet $\kappa$-(ET)$_2$Cu[N(CN)$_2$]Cl (N\'{e}el temperature $\sim$22~K \cite{Kagawa2008,Ito2015}) whose interlayer distance is close to the case of $\kappa$-(ET)$_2$Cu$_2$(CN)$_3$ \cite{Internal_field}. 
 We found the internal field at the $^{63}$Cu site to be $^{63}H_{\rm int}$ $\sim$ $70$~G \cite{Cu_NQR}. 
According to $^{13}$C-NMR measurement by Smith {\it et al}., the mean internal field at the $^{13}$C site is estimated to be $^{13}H_{\rm int}$ $\sim$ $750$~G \cite{Smith2003}. 
 Since the ratio of internal magnetic field ($|^{63}H_{\rm int}/^{13}H_{\rm int}| \sim 0.1$) is considered to be proportional to the hyperfine coupling constant ratio, $|^{63}A$/$^{13}A|^2$ is evaluated to be $0.01$, which is in good agreement with $(^{63}\gamma^{2}~^{63}T_1)^{-1}/(^{13}\gamma^{2}~^{13}T_1)^{-1} \sim 0.02$.
Therefore, the observed $^{63}T_{1}^{-1}$ between $10-40$~K can be interpreted as the magnetic fluctuations originated from the $\pi$ electrons on the ET layers.
The red solid curve shown in Fig.~\ref{fig3} is the sum of the two contributions of the magnetic relaxation ($^{63}T_1^{-1} \propto T^{1/2}$) and the quadruple relaxation $T_{1,{\rm Q}}^{-1}$: $T_1^{-1}$=$aT^{1/2}$ + $T_{1,{\rm Q}}^{-1}$ (with $a$=$0.055$), which reproduces the experimental data quite well.

%%%%%%%%%%%%%%%%%%%%%%%%%%%% FIG4 %%%%%%%%%%%%%%%%%%%%%%%%%%%%%%%%%%%
\begin{figure}[tbp]
\begin{center}
\includegraphics[width=\columnwidth]{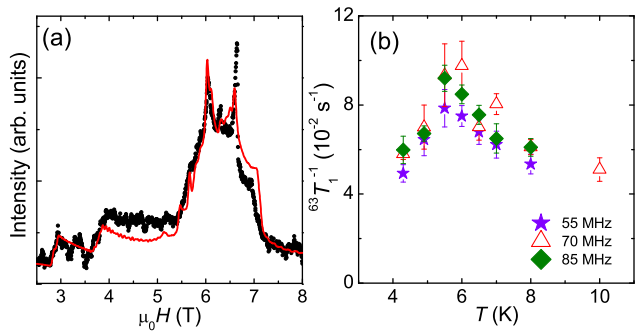}
\end{center}
\caption{
(a) Field-swept $^{63,65}$Cu-NMR spectra at $75$~MHz and $1.7$~K (black symbols) and the simulation result of four observed NQR frequency and $\eta$=$0.5$ (red line). 
(b) $T$ dependence of $^{63}T_1^{-1}$ in the magnetic fields. 
}
\label{fig4}
\end{figure}
%%%%%%%%%%%%%%%%%%%%%%%%%%%% FIG4 %%%%%%%%%%%%%%%%%%%%%%%%%%%%%%%%%%%

 The most striking feature in the $T$ dependence of $^{63}T_1^{-1}$ is the observation of the sharp peak around $6$~K, indicating a second-order phase transition.  
We exclude the possibility of crossover phenomenon for the anomaly by measuring the resonance frequency dependence of $^{63}T_1^{-1}$.
In the case of crossover, we expect Bloembergen-Purcell-Pound-like behavior where frequency-dependent $T_1^{-1}$ is expected \cite{Bloembergen1948}.
 Since one cannot change the resonance frequency in the NQR experiment, we have carried out NMR measurements. 
 Figure~\ref{fig4}(a) shows the field-swept $^{63,65}$Cu-NMR spectrum measured at a fixed frequency $75$~MHz at $1.7$~K where a broad and complicated spectrum is observed.
 This is due to large $\nu_{\rm NQR}$ and a finite value of $\eta$ as well as the superposition of four NMR lines ($^{63,65}$Cu ions for the two Cu sites). 
 The red curve is the sum of the four calculated powder-pattern spectra \cite{NMR_spectrum}. 
  As observed in the figure, the calculated spectrum well reproduces the characteristic shape of the observed spectrum. 
 $^{63}T_1$ was measured at the relatively sharp peak appeared at the lower magnetic field side of the main broad peak [around $6$~T for $75$~MHz in Fig.~\ref{fig4}(a)] while changing the frequency \cite{NMR_T1}.
  As shown in Fig.~\ref{fig4}(b), no obvious frequency dependence in $^{63}T_1$ was observed within our experimental uncertainty, evidencing again that the anomaly at $6$~K is not due to a crossover but to a phase transition with critical slowing down.

 What then is the origin of the phase transition?
  $^{13}T_1^{-1}$ measurements could not clearly detect the anomaly at $6$~K \cite{Shimizu2003,Shimizu2006}, suggesting that the $6$~K anomaly is not due to the magnetic fluctuation of the ET layers. 
  Furthermore, magnetic fluctuations originated from the anion insulating layers are also unlikely because Cu$^{+}$ ion is nonmagnetic and also $^{1}T_1^{-1}$ performed on the ethylene end groups of the ET molecules near the anion layer, showed no clear anomaly \cite{Shimizu2003}. 
 Therefore, it is concluded that the peak in the $T$ dependence of $^{63}T_{1}^{-1}$ originates from the EFG fluctuation. 
 There are three possible scenarios: the first one is the charge fluctuations due to the disorder of C and N ions in the anion layers, the second one is structural phase transition, and the last one is due to CD in the ET layers. 
 
 Regarding the first one, since two Cu sites, Cu1 and Cu2, caused by the disorder of C and N ions in the anion layers were observed below and above $6$~K, no change in the disorder is expected at $6$~K. 
 In addition, no change in $\nu_{\rm NQR}$ and linewidth at $6$~K also will exclude the possibility of structural phase transition at $6$~K. 

 Hence, the most probable explanation is the CD in the ET layers. 
 To check whether or not a CD on the conduction layers actually affects the EFG value at the Cu site, we have calculated the EFG while assuming a CD of $\pm 0.1e$ on the ET molecules.
 From our simple point charge calculation, we found that the EFG at the Cu sites is modulated by $0.5 \times 10^{18}$~V/m$^2$.
 The value is comparable to that from the thermal displacement of CN$^{-}$ ions which modulates the EFG by $2.2 \times 10^{18}$~V/m$^2$ at the Cu site \cite{Vibration}.  
 Since we actually observed the $T^{\sim2}$ dependence of $^{63}T_1^{-1}$ due to the thermal vibration of CN$^{-}$ ions at high $T$, we conclude that the critical slowing down of the CD on the ET molecules should be detected by $^{63}$Cu NQR at low $T$ when the lattice vibration is suppressed.
 
 According to the previous $^{13}$C-NMR studies, $^{13}$C-NMR linewidth discontinuously increases below $6$~K \cite{Kawamoto2004a,Shimizu2006}. 
 Since the linewidth relates to the distribution of charge density through the hyperfine couplings \cite{Kawamoto2004a,Shimizu2006}, the value of the discontinuous jump of the linewidth divided by the Knight shift (corresponding the mean value of charge density) will give an estimate of the degree of charge distribution. 
 Using the data from Ref. \cite{Kawamoto2004a}, we estimate the charge distribution from the average valence of the ET molecules to be ($0.5 \pm 0.13$)$e$ \cite{valence}. 
 The decrease in $\beta$ both in $^{13}$C NMR and $^{63}$Cu NQR can also be understood by the CD due to the phase transition at $6$~K. 
Therefore, it is quite reasonable that the charge degree of freedom in the ET layer is responsible for the observed phase transition. 
 CD within dimers has been proposed experimentally \cite{Abdel-Jawad2010,Itoh2013} and theoretically \cite{Naka2010,Hotta2010,Dayal2011,Gomi2013,Gomi2016}, and the recent structural analysis also pointed out the possibility of CD between dimers \cite{Foury-Leylekian2018}. 
On the other hand, powder transmission measurement detected no CD \cite{Sedlmeier2012}.
 Further studies are required to clarify what kind of CD occurs.

Finally we comment on small humps around $3$~K observed in $^{63}T_1^{-1}$ and $\beta$.  
The anomaly at the same temperature was also observed in such as thermal expansion, dielectric function, and $\mu$SR measurements, indicating it is intrinsic \cite{Manna2010,Poirier2012,Nakajima2012}.
At present,  we cannot conclude the origin of the hump behavior in $^{63}T_1^{-1}$ from our NQR experiments.  
Detailed low-$T$ Cu-NQR measurements down to such as $0.1$~K is interesting to shed the light on the physical properties of the compound at low $T$. 
This is a future work. 

 In conclusion, we performed the $^{63}$Cu-NQR measurement on the quantum spin-liquid candidate $\kappa$-(ET)$_2$Cu$_2$(CN)$_3$ to investigate the charge dynamics. 
Two different Cu sites are observed in the $^{63,65}$Cu NQR spectrum, which is direct evidence for the disorder of C and N atoms in the cyanide groups. 
$^{63}T_1^{-1}$ shows $T^2$ dependence indicating the EFG fluctuation due to the lattice vibration above $60$~K, below which $^{63}T_1^{-1}$ changes from $T^2$ to $T^{1/2}$ dependence with the suppression of lattice vibration. 
The $T^{1/2}$ dependence of $^{63}T_1^{-1}$ observed in $T$=$10-40$~K was ascribed to the magnetic fluctuation of $\pi$ electrons of the ET layers. 
Below $10$~K, $^{63}T_1^{-1}$ increases and divergent behavior was observed at $6$~K, evidencing a phase transition with critical slowing down derived from the EFG fluctuation. 
Based on our NQR data, we attributed the $6$~K anomaly to the phase transition of CD originated from the $\pi$ electrons in the ET layers. 
 Our results require to reconsider the current interpretation of the low-$T$ electronic state of $\kappa$-(ET)$_2$Cu$_2$(CN)$_3$, which was thought to exhibit a paramagnetic spin state without the phase transition down to a very low $T$ due to the spin frustration. 

\appendix
The authors are grateful to Y.~Saito, M.~Dressel, and M.~Lang for useful discussions. 
The research was supported by the U.S. Department of Energy (DOE), Office of Basic Energy Sciences, Division of Materials Sciences and Engineering. Ames Laboratory is operated for the U.S. DOE by Iowa State University under Contract No.~DE-AC02-07CH11358.
This work was partially supported by the Japan Society for the Promotion of Science KAKENHI Grant Numbers 18H05843, 19K21033, and 16K05427. 
T. K. also thanks the KAKENHI Grant Number JP16H01076: J-Physics for financial support to be a visiting scientist at the Ames Laboratory.

%\bibliography{aps}

\begin{thebibliography}{58}
\bibitem [{NMR({\natexlab{a}})}]{NMR_spectrum}%
  \BibitemOpen
  \href@noop {} {}\bibinfo {note} {NMR spectrum was
  calculated with a nuclear spin Hamiltonian including Zeeman and quadrupolar
  interactions without perturbation. In the calculation, we used the four
  values of $\nu_{\rm Q}$ determined by the NQR measurements and also $\eta$ =
  0.5 for each.}\BibitemShut {Stop}%
\bibitem [{NMR({\natexlab{b}})}]{NMR_T1}%
  \BibitemOpen
  \href@noop {} {} \bibinfo {note} {$T_1$ was estimated from
  the fitting with $1-M(t)/M(\infty) = 0.1\exp[-(t/T_1)^{\beta}] +
  0.9\exp[-(6t/T_1)^{\beta}]$ where the $\beta$ values change from 0.7 to 1.0
  depending on temperature and also resonance frequency.}\BibitemShut {Stop}%
\bibitem [{Vib()}]{Vibration}%
  \BibitemOpen
  \href@noop {} {}\bibinfo {note} {We assumed the amplitude of $5$~pm, which is
  estimated from $\sqrt{k_{\rm B}T/k}$, where \textit{k} is force constant.
  \textit{k} of ciano group can be estimated from the vaibrational mode of
  Cu-CN-Cu motion $\sim 500$~cm$^{-1}$ [10].}\BibitemShut {Stop}%
\bibitem [{val()}]{valence}%
  \BibitemOpen
  \href@noop {} {}\bibinfo {note} {From Figs. 3 and 4 in Ref.~[26], Knight shift
  and linewidth jump at $6$~K are $270$~ppm and $7$~kHz (corresponding to
  $70$~ppm), respectively, which give the fractional charge ratio of $\sim
  0.26$.}\BibitemShut {Stop}%
\end{thebibliography}
%

\end{document}